\documentclass[3p,preprint]{elsarticle}
\usepackage[utf8]{inputenc}
\usepackage{amsmath}
\usepackage{amssymb}
\usepackage{amsfonts}
\usepackage{graphicx}
\usepackage{epstopdf}
\usepackage{psfrag}
\usepackage[hidelinks]{hyperref}
\usepackage{doi}
\usepackage{xcolor}
\usepackage{cancel}
\hypersetup{
    colorlinks,
    linkcolor={blue},
    citecolor={blue},
    urlcolor={blue}
}

\begin{document}

\title{Stability of gap soliton complexes in the nonlinear Schr$\ddot{\rm o}$dinger equation with periodic potential and repulsive nonlinearity}
\author{P. P. Kizin\corref{ppk}}
\cortext[ppk]{e-mail: p.p.kizin@gmail.com}

\address{Moscow Institute of Electronic Engineering, 
Zelenograd, Moscow, 124498, Russia}

\begin{abstract}
The work is devoted to numerical investigation of stability of stationary localized modes (``gap solitons'') for the one-dimentional nonlinear Schr$\ddot{\rm o}$dinger equation (NLSE) with periodic potential and repulsive nonlinearity. Two classes of the modes are considered: a bound state of a pair of {\it in-phase} and {\it out-of-phase} fundamental gap solitons (FGSs) from the first bandgap  separated by various number of empty potential wells. Using the standard framework of linear stability analysis, we computed the linear spectra for the gap solitons by means of the Fourier collocation method and the Evans function method. We found that the gap solitons of the first and second classes are exponentially unstable for  odd and even numbers of separating periods of the potential, respectively. The real parts of unstable eigenvalues in corresponding spectra decay with the distance between FGSs exponentially. On the contrary, we observed that the modes of the first and second classes are either linearly stable or exhibit weak oscillatory instabilities if the number of empty potential wells separating FGSs is even and odd, respectively. In both cases, the oscillatory instabilities arise in some vicinity of upper bandgap edge. In order to check the linear stability results, we fulfilled numerical simulations for the time-dependent NLSE by means of a finite-difference scheme. As a result, all the considered exponentially unstable solutions have been deformed to long-lived pulsating formations whereas stable solutions conserved their shapes for a long time.
\end{abstract}

\begin{keyword}
Nonlinear Schr$\ddot{\rm o}$dinger equation, 
periodic potential, 
gap solitons,
stability
\end{keyword}

\maketitle

\section{Introduction} \label{Intro}

In the past few decades the nonlinear Schr$\ddot{\rm o}$dinger equation (NLSE) with an additional non-autonomous linear term has become one of the actual and challenging physical problems which has been studied worldwide by different scientific groups. This equation can be written in the following three-dimensional form:
\begin{equation}\label{eq:nls}
iU_t=-\Delta U +V(x,y,z)U + \sigma |U|^2 U,
\end{equation}
\begin{equation*}
U=U(x,y,z,t), \quad
\Delta=\frac{\partial^2}{\partial x^2}+\frac{\partial^2}{\partial y^2}+\frac{\partial^2}{\partial z^2}, \quad
\sigma=\pm1.
\label{eq:Delta}
\end{equation*}
From a physical viewpoint, equation~(\ref{eq:nls}) is related to the models of
plasma physics \cite{berge,chen}, nonlinear optics \cite{edwards,kunze,kumar,turitsyn_2,turitsyn_1} and Bose-Einstein condensation theory \cite{dalfovo}. It is worth mentioning that in the latter context, equation~(\ref{eq:nls}), also called the Gross-Pitaevskii equation (GPE), has become especially relevant after experimental observation of the Bose-Einstein condensate in 1995 \cite{anderson,bradley,davis}. Such a state of matter has been predicted in 1924 by Einstein and Bose \cite{bose,einstein}. In the context of the meanfield theory of the Bose-Einstein condensate, the term $|U|^2$ describes the local density of the condensate. The function $V(x,y,z)$ has the meaning of an external potential which allows one to confine the condensate spatially. Parameter $\sigma$ describes interparticle interactions in the condensate: the value $\sigma=+1$ corresponds to repulsive interactions between atoms whereas $\sigma=-1$ corresponds to attractive interactions.

\begin{figure}\centering
\begin{psfrags}
\psfrag{A}[bl][bl][1.5]{$V_0$}
\psfrag{B}[bl][bl][1.5]{$\mu$}
\psfrag{C}[bl][bl][1]{$0$}
\psfrag{D}[bl][bl][1]{$2$}
\psfrag{E}[bl][bl][1]{$4$}
\psfrag{F}[bl][bl][1]{$6$}
\psfrag{G}[bl][bl][1]{$8$}
\psfrag{H}[bl][bl][1]{$10$}
\psfrag{I}[bl][bl][1]{$12$}
\psfrag{J}[bl][bl][1]{$-6$}
\psfrag{K}[bl][bl][1]{$-4$}
\psfrag{L}[bl][bl][1]{$-2$}
\psfrag{M}[bl][bl][1]{$0$}
\psfrag{N}[bl][bl][1]{$2$}
\psfrag{O}[bl][bl][1]{$4$}
\psfrag{P}[bl][bl][1]{$6$}
\psfrag{Q}[bl][bl][1]{$8$}
\psfrag{R}[bl][bl][1.2]{Region-1}
\psfrag{S}[bl][bl][1.2]{Region-2}
\psfrag{T}[bl][bl][1.2]{Gap-1}
\psfrag{U}[bl][bl][1.2]{Gap-2}
\includegraphics[width=0.5\textwidth]{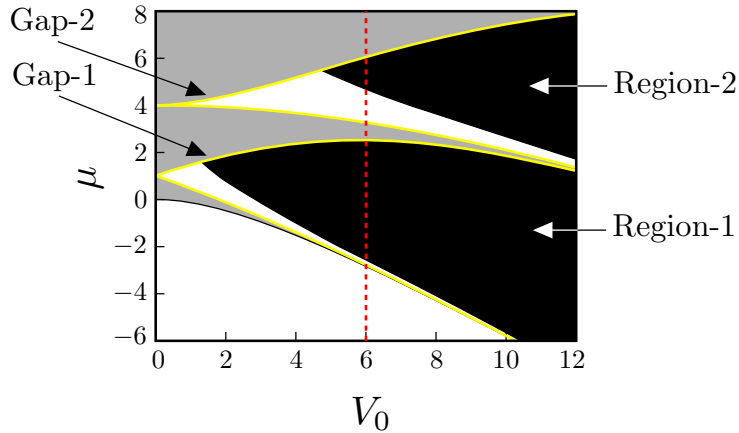}
\caption{A band-gap structure of linear equation (\ref{eq:linear}). Gray areas correspond to the bands of the spectrum $\mu$. White and black areas represent the bandgaps. In addition, black regions depict a parametric area where all solutions of (\ref{eq:main}) can be coded by symbolic sequences. Red dashed line represents the value $V_0$ considered in the present study.}
\label{fig:bandgap}
\end{psfrags}
\end{figure}

The one-dimensional version of (\ref{eq:nls}):
\begin{equation}\label{eq:1d}
iU_t = -U_{xx}+V(x)U+\sigma|U|^2 U,
\end{equation}
\begin{equation*}
U = U(x,t), \quad \sigma=\pm1,
\end{equation*}
describes an elongated (cigar-shaped) cloud of the condensate.
An important set of solutions of (\ref{eq:1d}) is a class of so-called stationary modes having the following form:
\begin{equation}\label{eq:anzatz}
U(x,t)=u(x)e^{-i\mu t},
\end{equation}
where $\mu$ has the meaning of the chemical potential of the condensate. Stationary modes $u(x)$ satisfy the localization condition:
\begin{equation}\label{eq:local}
\lim_{x\to\pm\infty}{u(x)}=0.
\end{equation}
In what follows, it is assumed that $u(x)$ is a real-valued function \cite{alfimov_3}. Besides, the parameter $\sigma$  is chosen according to the repulsive case of interactions, $\sigma=+1$, and the potential $V(x)$ is fixed in the model form $V(x)=-V_0\cos{2x}$. Here, the parameter $V_0$ is related to the depth of potential wells. Substituting (\ref{eq:anzatz}) into (\ref{eq:1d}), one arrives at the following time-independent equation:
\begin{equation}\label{eq:main}
\mu u = -u_{xx}-V_0\cos(2x)u+u^3.
\end{equation}
Due to the localization condition (\ref{eq:local}), for $|x|\gg 1$ equation (\ref{eq:main}) can be replaced by the linear Schr$\ddot{\rm o}$dinger equation:
\begin{equation}\label{eq:linear}
\mu u = -u_{xx}-V_0\cos(2x)u.
\end{equation}
It is well-known \cite{berezin} that the spectrum of problem (\ref{eq:linear}) has a band-gap structure which is depicted in Figure~\ref{fig:bandgap}. There is a countable number of bands separated by bandgaps. Such a band-gap structure of linearized equation imposes several restrictions for existence of localized solutions of (\ref{eq:main}). If the point $(\mu,V_0)$ lies in a band, equation  (\ref{eq:linear}) has no solutions that tend to zero at $+\infty$ or at $-\infty$. It means that  the localization condition (\ref{eq:local}) for equation (\ref{eq:main}) for these values $\mu$ and $V_0$ cannot be satisfied, except for the trivial solution $u(x)\equiv 0$. Therefore, the stationary localized modes (\ref{eq:anzatz}) can be obtained only in bandgaps. For this reason, the localized solutions of (\ref{eq:main}) are also called {\it gap solitons}.

The simplest localized solution of (5) is called \textit{fundamental gap soliton} (FGS). This solution represents a single density hump with decaying tails. It was proven in \cite{alfimov_2} that more complex solitons can be considered as compositions of several FGSs situated at different potential wells. More precisely,   all solutions of (5) situated in Region 1 in Figure 1 can be coded by means of bi-infinite symbolic sequences of the finite alphabet $\mathcal{A}=\{+,0,-\}$. The symbol ``$+$'' (or ``$-$'') at the $n$-th entry of the bi-infinite sequence ($n$ runs over the all integers from $-\infty$ to $+\infty$) indicates that the $n$-th potential well is occupied by the FGS (or by the FGS taken with negative sign). Respectively, the zero symbol means that the corresponding potential well is "empty". Thus, the code of the FGS has the form $(\ldots, 0, 0, +,  0, 0, \ldots)$, and the code of the FGS taken with negative sign has the form $(\ldots, 0, 0, -,  0, 0, \ldots)$. More generally, the code of an arbitrary gap soliton ``starts'' and ``ends'' with the infinite number of zero symbols, i.e. has the form $(\ldots,0,0,s_1,s_2,\ldots,s_n,0,0,\ldots)$, $0\in\mathcal{A}$, $s_i\in\mathcal{A}$, $s_1\neq 0$, $s_n\neq 0$. These infinite sequences of zero symbols describe the asymptotically decaying tails of the soliton. For the sake of brevity, in what follows we omit  the zero symbols situated at the soliton tails. Then the code of the FGS can be written as $(+)$. The code $(++)$ corresponds to a soliton composed of two neighbor FGSs,  both  taken with positive sign,  and the code $(+-)$ describes a composition of two FGSs taken with different signs. The code $(+0+)$ describes a bound state of two FGSs separated by an ``empty'' potential well.

An important property of a gap soliton is its stability which indicates the robustness of the soliton against perturbations. Various articles have reported analytical and numerical results about stability of gap solitons of (\ref{eq:main}). Let us list the known results about stability of gap solitons of (\ref{eq:main}) situated in Region~1 in Figure~\ref{fig:bandgap}.

In was claimed in \cite{pelinovsky} that the small-amplitude FGS having the code $(+)$ is linearly stable. However, in \cite{pelinovsky} it was also  suggested that the FGS of greater amplitudes may suffer weak {\it oscillatory instabilities} caused by complex eigenvalues in the linear spectrum [see the spectral problem (\ref{eq:spectral})]. These instabilities were found in the same paper  for  the case of attractive nonlinearity [i.e., the case $\sigma=-1$ in (\ref{eq:1d})]. Turning to the large amplitudes, the stability of the FGS was examined numerically in \cite{louis} where the  authors concluded that it is a stable mode as well.  Besides, the authors of \cite{louis} mentioned  their finding of complex eigenvalues with a small non-zero real parts for (\ref{eq:spectral}), but they attributed them to the insufficient accuracy of their computations. Recently, in \cite{kizin} the stability of the FGS  was studied using the Evans function approach, and it was concluded that the spectrum of the FGS does include complex eigenvalues with small real parts which leads to weak instabilities. Therefore, the FGS typically is either stable or exhibits weak oscillatory instabilities.

The next basic solution of (\ref{eq:main}) has the code $(+-)$. It was studied analitically in \cite{hwang,pelinovsky} using the small-amplitude approximation and turned out to be unstable (the spectrum includes a pair of real eigenvalues which implies so-called {\it exponential instability}). For large amplitudes,  this solution was studied numerically \cite{louis} and turned out to be exponentially unstable as well.

The stability of more complex modes of (\ref{eq:main}), double-humped and triple-humped gap solitons with the codes $(++)$ and $(+++)$ (also called \textit{Truncated-Bloch-wave solitons}), was checked numerically in \cite{wang} where these solutions were reported to be stable. However, it was shown recently in \cite{kizin} that these solutions also may suffer from small oscillatory instabilities.

Two families of separated FGSs having the codes $(+0+)$ and $(+0-)$ have been studied numerically in \cite{louis} where they were found to be oscillatory unstable. However, the stability of the gap solitons with codes $(+0\ldots0+)$ and $(+0\ldots0-)$, i.e. having the symbols ``$+$'' or ``$-$'' separated by some number of symbols ``$0$'', up to the moment have not been studied in detail. An interesting analysis of  bifurcations of these solutions can be found in \cite{akylas}. In the present study we focus our attention on compexes of gap solitons of (\ref{eq:main}) having the codes $(+0\ldots0+)$ and $(+0\ldots0-)$ with different number of separating periods. We do not consider the codes $(-0\ldots0-)$ and $(-0\ldots0+)$ due to the symmetry of (\ref{eq:main}). We study numerically the linear stability of these modes and check it by direct integration of equation~(\ref{eq:1d}) using an appropriate numerical scheme. In Section~1 we describe the linear stability problem in detail and describe the required numerical methods. In Section~2 we present the main outcomes of this paper and in Section~3 we summarize the results of the work.

\section{Stability of gap solitons}
\label{sec:stab}

\begin{figure}\centering
\begin{psfrags}
\psfrag{A}[bl][bl][1.5]{$x$}
\psfrag{B}[bl][bl][1.5]{$x$}
\psfrag{C}[bl][bl][1.5]{$x$}
\psfrag{D}[bl][bl][1.5]{$u$}
\psfrag{E}[bl][bl][1.5]{$u$}
\psfrag{F}[bl][bl][1]{-$3$}
\psfrag{G}[bl][bl][1]{-$2$}
\psfrag{H}[bl][bl][1]{-$1$}
\psfrag{I}[bl][bl][1]{$0$}
\psfrag{J}[bl][bl][1]{$1$}
\psfrag{K}[bl][bl][1]{$2$}
\psfrag{L}[bl][bl][1]{$3$}
\psfrag{M}[bl][bl][1]{$4$}
\psfrag{N}[bl][bl][1]{$5$}
\psfrag{O}[bl][bl][1]{-$3$}
\psfrag{P}[bl][bl][1]{-$2$}
\psfrag{Q}[bl][bl][1]{-$1$}
\psfrag{R}[bl][bl][1]{$0$}
\psfrag{S}[bl][bl][1]{$1$}
\psfrag{T}[bl][bl][1]{$2$}
\psfrag{U}[bl][bl][1]{$3$}
\psfrag{V}[bl][bl][1]{$4$}
\psfrag{W}[bl][bl][1]{$5$}
\psfrag{X}[bl][bl][1]{-$2$}
\psfrag{Y}[bl][bl][1]{-$1$}
\psfrag{Z}[bl][bl][1]{$0$}
\psfrag{a}[bl][bl][1]{$1$}
\psfrag{b}[bl][bl][1]{$2$}
\psfrag{c}[bl][bl][1]{$3$}
\psfrag{d}[bl][bl][1]{$4$}
\psfrag{e}[bl][bl][1]{$5$}
\psfrag{f}[bl][bl][1]{$6$}
\psfrag{g}[bl][bl][1]{-$2$}
\psfrag{h}[bl][bl][1]{-$1$}
\psfrag{i}[bl][bl][1]{$0$}
\psfrag{j}[bl][bl][1]{$1$}
\psfrag{k}[bl][bl][1]{$2$}
\psfrag{l}[bl][bl][1]{-$2$}
\psfrag{m}[bl][bl][1]{-$1$}
\psfrag{n}[bl][bl][1]{$0$}
\psfrag{o}[bl][bl][1]{$1$}
\psfrag{p}[bl][bl][1]{$2$}
\psfrag{q}[bl][bl][1.2]{(a)}
\psfrag{r}[bl][bl][1.2]{(b)}
\psfrag{s}[bl][bl][1.2]{(c)}
\psfrag{t}[bl][bl][1.2]{(d)}
\psfrag{u}[bl][bl][1.2]{(e)}
\psfrag{v}[bl][bl][1.2]{(f)}
\psfrag{w}[bl][bl][1]{$\times\pi$}
\psfrag{x}[bl][bl][1]{$\times\pi$}
\psfrag{y}[bl][bl][1]{$\times\pi$}
\includegraphics[width=0.75\textwidth]{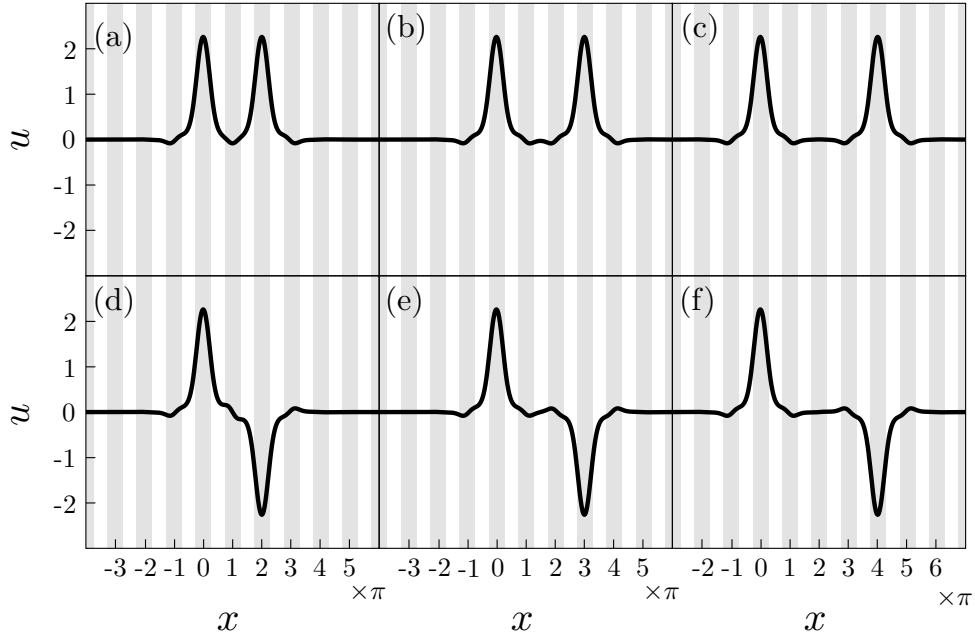}
\caption{Two classes of gap solitons of (\ref{eq:main}) considered in the present study. The first class consists of two separated in-phase FGSs with the codes of the form $(+0\ldots0+)$: (a) $(+0+)$, (b) $(+00+)$, (c) $(+000+)$. The second class consists of two separated out-of-phase FGSs with the codes $(+0\ldots0-)$: (d) $(+0-)$, (e) $(+00-)$, (f) $(+000-)$. The parameters are $\mu=1$ and $V_0=6$.}
\label{fig:exam}
\end{psfrags}
\end{figure}

\subsection{Linear stability problem}

From a physical viewpoint, the stability is an important property of solutions of equation~(\ref{eq:1d}). We study the linear (spectral) stability following traditional approach \cite{yang}. Let $u(x)$ be an arbitrary solution of (\ref{eq:main}) satisfying the boundary conditions (\ref{eq:local}). This solution can be regarded as the initial condition $U(x,0)$ for equation (\ref{eq:1d}). Adding a small perturbation to the function $u(x)$ and substituting it to (\ref{eq:anzatz}), one has the following perturbed solution of (\ref{eq:1d}):
\begin{equation}\label{eq:pertrub}
U(x,t)=\left(u(x)+e^{\lambda t}[a(x)+ib(x)]\right)e^{-i\mu t},
\end{equation}
where $a(x)$ and $b(x)$ are real and small enough, $|a|,|b|\ll 1$. Substituting (\ref{eq:pertrub}) into (\ref{eq:1d}) and omitting all high-order terms with respect to $a(x)$ and $b(x)$, we come to the following system:
\begin{equation}\label{eq:syst}
\left\{
\begin{array}{l}
\lambda b = a_{xx}+\left(\mu-V(x)\right)a-3u^2a, \\
\lambda a = -b_{xx}-\left(\mu-V(x)\right)b+u^2b.
\end{array}
\right.
\end{equation}
Introducing the linear operators $\mathcal{L}_0=d^2/dx^2+(\mu-V(x))-u^2$ and $\mathcal{L}_1=d^2/dx^2+(\mu-V(x))-3u^2$, the system (\ref{eq:syst}) can be rewritten in the form of the spectral problem:
\begin{equation}\label{eq:spectral}
\lambda\left(
\begin{array}{l}
a \\ b
\end{array}
\right)
=
\left(
\begin{array}{cc}
0 & -\mathcal{L}_0 \\
\mathcal{L}_1 & 0
\end{array}
\right) 
\left(
\begin{array}{l}
a \\ b
\end{array}
\right).
\end{equation}
Generally, the spectrum of (\ref{eq:spectral}) consists of an essential and a discrete parts. Since the linear stability of considered modes is related to the latter part, let us mention some properties of discrete eigenvalues of (\ref{eq:spectral}). The eigenvalue $\lambda=0$ is double, with eigenvector $(0, u)^T$ and generalized eigenvector $(\partial{u}/\partial\mu, 0)^T$ that obey:
\begin{equation}
0=
\left(
\begin{array}{cc}
0 & -\mathcal{L}_0 \\
\mathcal{L}_1 & 0
\end{array}
\right) 
\left(
\begin{array}{l}
0 \\ u
\end{array}
\right),
\quad
0=
\left(
\begin{array}{cc}
0 & -\mathcal{L}_0 \\
\mathcal{L}_1 & 0
\end{array}
\right)^2
\left(
\begin{array}{c}
\partial{u}/\partial\mu \\ 0
\end{array}
\right),
\end{equation}
respectively. This fact allows one to check the accuracy of numerical techniques in a simple way. We use it to check the Evans function method which is described in \cite{blank} in detail. Besides, if the spectrum of (\ref{eq:spectral}) contains an isolated eigenvalue $\lambda$, it also contains the values $-\lambda$, $\lambda^*$ and $-\lambda^*$ where the asterisk means the complex conjugation.

Solving the problem (\ref{eq:spectral}) one concludes whether the regarding mode (\ref{eq:anzatz}) is stable or unstable. In the case the whole spectrum is pure imaginary, the solution (\ref{eq:anzatz}) is said to be linearly stable. On the other hand, if there is at least one eigenvalue with non-zero real part, then the perturbation grows exponentially and the solution (\ref{eq:anzatz}) is regarded to be unstable.

\subsection{Numerical methods}

In order to construct the gap solitons of (\ref{eq:main}), we applied the modified shooting algorithm \cite{alfimov_4}. This algorithm allows one to compute the profile of a required mode by its symbolic representation. It includes solving of a Cauchy problem for equation~(\ref{eq:main}) which was implemented using the Runge-Kutta 4-order scheme.

Concerning the numerical solution of the spectral problem (\ref{eq:spectral}), we used two alternative algorithms. One of them is the well-elaborated Fourier collocation method (FCM, see \cite{yang} for details). It is well-suited for fast and accurate detecting of relatively strong instabilities. Hovewer, its accuracy may not be sufficient for tracing relatively weak instabilities associated with eigenvalues with small nonzero real parts. For accurate detection of such weak oscillatory instabilities, we  take advantage of the second algorithm based on the Evans function (EFM) that allows one to compute complex eigenvalues with tiny real parts with high precision. On the other hand, the realization of EFM is quite sophisticated (see \cite{blank}).

\begin{figure}\centering
\begin{psfrags}
\psfrag{A}[bl][bl][1.2]{${\rm Re}\,\lambda$}
\psfrag{B}[bl][bl][1.2]{${\rm Re}\,\lambda$}
\psfrag{C}[bl][bl][1.2]{${\rm Re}\,\lambda$}
\psfrag{D}[bl][bl][1.2]{${\rm Im}\,\lambda$}
\psfrag{E}[bl][bl][1.2]{${\rm Im}\,\lambda$}
\psfrag{F}[bl][bl][1]{$-0.2$}
\psfrag{G}[bl][bl][1]{$0$}
\psfrag{H}[bl][bl][1]{$0.2$}
\psfrag{I}[bl][bl][1]{$-0.05$}
\psfrag{J}[bl][bl][1]{$0$}
\psfrag{K}[bl][bl][1]{$0.05$}
\psfrag{L}[bl][bl][1]{$-0.01$}
\psfrag{M}[bl][bl][1]{$0$}
\psfrag{N}[bl][bl][1]{$0.01$}
\psfrag{O}[bl][bl][1]{$-10$}
\psfrag{P}[bl][bl][1]{$-5$}
\psfrag{Q}[bl][bl][1]{$0$}
\psfrag{R}[bl][bl][1]{$5$}
\psfrag{S}[bl][bl][1]{$10$}
\psfrag{T}[bl][bl][1]{$-10$}
\psfrag{U}[bl][bl][1]{$-5$}
\psfrag{V}[bl][bl][1]{$0$}
\psfrag{W}[bl][bl][1]{$5$}
\psfrag{X}[bl][bl][1]{$10$}
\psfrag{Y}[bl][bl][1.2]{(a)}
\psfrag{Z}[bl][bl][1.2]{(b)}
\psfrag{a}[bl][bl][1.2]{(c)}
\psfrag{b}[bl][bl][1.2]{(d)}
\psfrag{c}[bl][bl][1.2]{(e)}
\psfrag{d}[bl][bl][1.2]{(f)}
\includegraphics[width=0.75\textwidth]{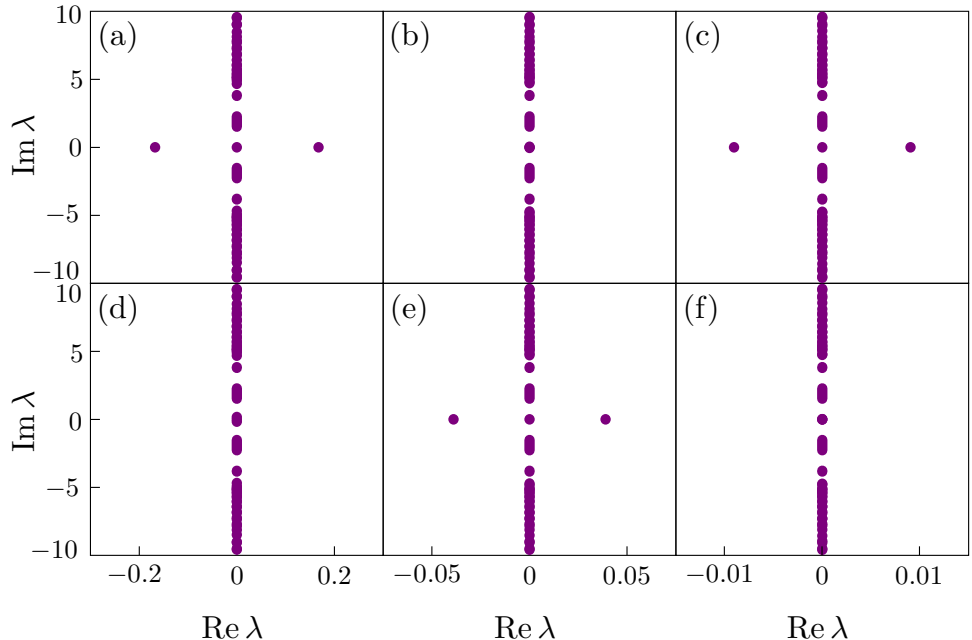}
\caption{Eigenvalues (\ref{eq:spectral}) of gap solitons depicted in Figure~\ref{fig:exam}.  The parameters are $\mu=1$, $V_0=6$.}
\label{fig:spectrum}
\end{psfrags}
\end{figure}

\begin{figure}\centering
\begin{psfrags}
\psfrag{A}[bl][bl][1.2]{Number of empty sites}
\psfrag{R}[bl][bl][1.2]{between FGSs}
\psfrag{B}[bl][bl][1.5]{$|{\rm Re}\,\lambda|$}
\psfrag{C}[bl][bl][1]{$1$}
\psfrag{D}[bl][bl][1]{$2$}
\psfrag{E}[bl][bl][1]{$3$}
\psfrag{F}[bl][bl][1]{$4$}
\psfrag{G}[bl][bl][1]{$5$}
\psfrag{H}[bl][bl][1]{$10^{-4}$}
\psfrag{I}[bl][bl][1]{$10^{-3}$}
\psfrag{J}[bl][bl][1]{$10^{-2}$}
\psfrag{K}[bl][bl][1]{$10^{-1}$}
\psfrag{L}[bl][bl][1]{$1$}
\psfrag{M}[bl][bl][1]{$(+0+)$}
\psfrag{N}[bl][bl][1]{$(+00-)$}
\psfrag{O}[bl][bl][1]{$(+000+)$}
\psfrag{P}[bl][bl][1]{$(+0000-)$}
\psfrag{Q}[bl][bl][1]{$(+00000+)$}
\includegraphics[width=0.38\textwidth]{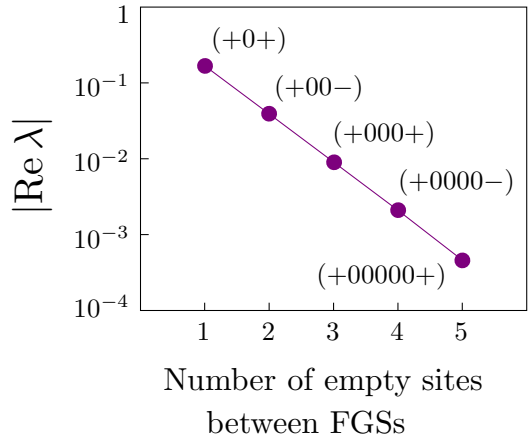}
\caption{Real eigenvalues in the spectrum (\ref{eq:spectral}) for different gap solitons. The codes of considered gap solitons are given beside the points. The parameters are $\mu=1$, $V_0=6$.}
\label{fig:distance}
\end{psfrags}
\end{figure}

The linear stability results for gap solitons of (\ref{eq:main}) obtained by both FCM and EFM allow us to predict the dynamics of corresponding solution of (\ref{eq:1d}) in time. To check such a prediction, we integrated (\ref{eq:1d}) by means of a finite difference scheme described in \cite{trofimov}. In the case the solution $U(x,t)$ undergoes any type of instabilities, the profile of initial condition $U(x,0)$ is expected to undergo deformation due to the growth of the perturbation whereas the profile of a stable solution is expected to approximately preserve its shape for any period of time.

\section{Results}
\label{sec:res}

Using the numerical schemes described above we fulfilled two series of numerical experiments for  equations~(\ref{eq:1d}) and (\ref{eq:spectral}). The Sections~2.1 and 2.2 below are devoted to two classes of gap solitons consisting of two separated in-phase and out-of-phase FGSs, respectively. Notice that since the profile of each  gap soliton depends on the parameters $\mu$ and $V_0$, for a fixed $V_0$ there is one-parametric family of gap solitons having the same code.
Since only two gap solitons with the codes $(+)$ and $(+-)$ can be continued to the lower boundary of the first bandgap, $\mu=\mu_-$, the solutions with all other codes  ``die'' in bifurcations for some greater values $\mu>\mu_-$.

\subsection{The stability of in-phase FGSs}

We started our investigation with bound states consisting of two in-phase FGSs separated by an arbitrary number of empty potential wells. Such solutions have the codes of the form $(+0\ldots0+)$ and $(-0\ldots0-)$ with various numbers of zero symbols ``0'' in the middle. The simplest representatives of this class are depicted in Figure~\ref{fig:exam}(a)-(c) for parameter $\mu=1$. On the one hand, they all are situated in Region 1 in Figure~\ref{fig:bandgap}, far away from the upper bandgap edge, which implies the absence of weak oscillatory instabilities in their spectra \cite{kizin}. On the other hand, they feature different stability properties. Looking at corresponding spectra presented in Figure~\ref{fig:spectrum}(a)-(c), one can conclude that the modes (a) and (c) are stable whereas the solution (b) undergoes a strong exponential instability caused by a pair of real eigenvalues.

\begin{figure}\centering
\begin{psfrags}
\psfrag{A}[bl][bl][1.5]{$\mu$}
\psfrag{B}[bl][bl][1.5]{$\mu$}
\psfrag{C}[bl][bl][1.5]{$M$}
\psfrag{D}[bl][bl][1.5]{$M$}
\psfrag{E}[bl][bl][1]{$-3$}
\psfrag{F}[bl][bl][1]{$-2$}
\psfrag{G}[bl][bl][1]{$-1$}
\psfrag{H}[bl][bl][1]{$0$}
\psfrag{I}[bl][bl][1]{$1$}
\psfrag{J}[bl][bl][1]{$2$}
\psfrag{K}[bl][bl][1]{$3$}
\psfrag{L}[bl][bl][1]{$-3$}
\psfrag{M}[bl][bl][1]{$-2$}
\psfrag{N}[bl][bl][1]{$-1$}
\psfrag{O}[bl][bl][1]{$0$}
\psfrag{P}[bl][bl][1]{$1$}
\psfrag{Q}[bl][bl][1]{$2$}
\psfrag{R}[bl][bl][1]{$3$}
\psfrag{S}[bl][bl][1]{$0$}
\psfrag{T}[bl][bl][1]{$40$}
\psfrag{U}[bl][bl][1]{$120$}
\psfrag{V}[bl][bl][1]{$200$}
\psfrag{W}[bl][bl][1]{$0$}
\psfrag{X}[bl][bl][1]{$40$}
\psfrag{Y}[bl][bl][1]{$120$}
\psfrag{Z}[bl][bl][1]{$200$}
\psfrag{a}[bl][bl][1.2]{(a)}
\psfrag{b}[bl][bl][1.2]{(b)}
\psfrag{c}[bl][bl][1.2]{Gap-1}
\psfrag{d}[bl][bl][1.2]{Gap-1}
\psfrag{e}[bl][bl][0.8]{$-2.79$}
\psfrag{f}[bl][bl][0.8]{$-2.73$}
\psfrag{g}[bl][bl][0.8]{$0.2$}
\psfrag{h}[bl][bl][0.8]{$0.8$}
\psfrag{i}[bl][bl][0.8]{$1.4$}
\psfrag{j}[bl][bl][0.8]{$-2.79$}
\psfrag{k}[bl][bl][0.8]{$-2.73$}
\psfrag{l}[bl][bl][0.8]{$0.2$}
\psfrag{m}[bl][bl][0.8]{$0.8$}
\psfrag{n}[bl][bl][0.8]{$1.4$}
\psfrag{o}[bl][bl][0.8][48]{$+0+$}
\psfrag{p}[bl][bl][0.8][62]{$+00+$}
\psfrag{q}[bl][bl][0.8][75]{$+000+$}
\psfrag{r}[bl][bl][0.8][48]{$+0-$}
\psfrag{s}[bl][bl][0.8][62]{$+00-$}
\psfrag{t}[bl][bl][0.8][75]{$+000-$}
\includegraphics[width=0.75\textwidth]{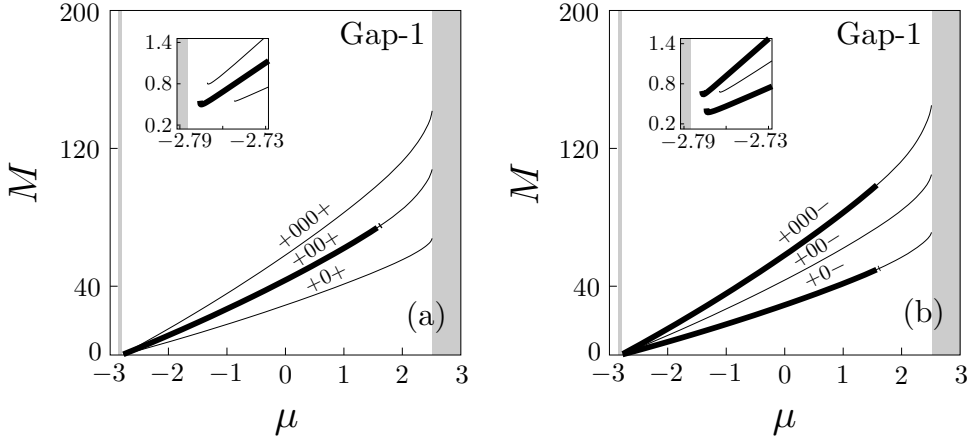}
\caption{Families of the gap solitons depicted in Figure~\ref{fig:exam}. Panel (a): families $(\mu,M)$ of two separated in-phase FGSs. Panel (b): families $(\mu,M)$ of two separated out-of-phase FGSs. Thin lines correspond to linearly unstable solutions while the bold lines represent stable ones. The left edges of presented families are depicted in the insets.}
\label{fig:families}
\end{psfrags}
\end{figure}

Let us illustrate the results by fixing the depth of the potential $V_0=6$. The lower boundary of the first bandgap in this case  is $\mu_-\approx-2.785$. Being continued from the upper edge of the bandgap, the solutions with the codes $(+0\ldots0+)$ and $(-0\ldots0-)$ ``die'' in bifurcations at some values of $\mu>\mu_-$. For instance, the gap solitons (a)-(c) bifurcate at $\mu\approx-2.752$, $\mu\approx-2.775$ and $\mu\approx-2.771$, respectively. Traditionally, a family of gap solitons is described by a curve on the plane $(\mu,N)$ where $N$ is squared $L^2$-norm, $N=\displaystyle{\int_\mathbb{R}}{u^2\,dx}$, of a gap soliton. However, for the solutions formed by a pair of well-separated FGSs such representation is not convenient, since their  $L^2$-norms are very close to each other and are equal approximately to the doubled $L^2$-norm of FGS [for instance, the families of gap solitons (a)-(c)  are almost indistinguishable on the plane $(\mu,N)$]. Therefore we choose another characteristic, which is free of this disadvantage and allows us to display the families properly:
\begin{equation}\label{eq:m}
M=\int_{\mathbb{R}}xu^2\,dx.
\end{equation}
Notice that if a solution $u(x)$ is either even or odd then its characteristic $M$ is equal to zero. Therefore we shift the solutions shown in Figure~\ref{fig:exam} in such a way that one of two FGSs is situated in zero potential well, $x=0$. The corresponding families on the plane $(\mu,M)$ are depicted in Figure~\ref{fig:families}(a) by black curves. Thin and thick segments of a curve are related to regions of linear stability and instability of gap solitons, respectively. These curves are situated in the white spacious region of Figure~\ref{fig:families} which represents the first bandgap. We found that the whole families (a) and (c) are linearly unstable due to the presence of one pair of  real eigenvalues in spectrum of (\ref{eq:spectral}). In the same time, the family (b) is stable for $\mu_-<\mu<\mu^*$ where the threshold value is $\mu^*\approx1.57$. If $\mu$ is greater than $\mu^*$, the family (b) undergoes weak oscillatory instabilities which are similar to those described in \cite{kizin}. For instanse, the spectrum of solution (b) includes the quartet of complex eigenvalues with real part ${\rm Re}\,\lambda\sim10^{-5}$ in the case $\mu=1.7$ and ${\rm Re}\,\lambda\sim10^{-4}$ as $\mu=2.3$. In addition, the inset presented in the left top corner in Figure~\ref{fig:families}(a) zooms in the bifurcation edges of considered families. 

To summarize, our numerical observations allow one to conjecture that {\it if the number of empty potential wells between two in-phase FGSs (zero symbols in middle of the code)  is odd, the composed gap soliton is unstable due to a strong exponential instability. Otherwise, it is  either stable or exhibits weak oscillatory instability.}

\subsection{The stability of out-of-phase FGSs}

Now let us consider the second class of bound states consistsing of two separated out-of-phase FGSs with the codes of the form $(+0\ldots0-)$ and $(-0\ldots0+)$. Let us fix again the depth of the potential, $V_0=6$. One can find the simplest states from this class in Figure~\ref{fig:exam}(d)-(f) for parameter $\mu=1$. Using FCM and EFM, we found that the gap soliton (d) and (f) are linearly stable, see Figure~\ref{fig:spectrum}(d), (f). On the contrary, the gap soliton (e) was found to be unstable, see Figure~\ref{fig:spectrum}(e).

The families of bound states (d)-(f), see Figure~\ref{fig:families}(b), can be extended numerically from the value $\mu=1$ to the smaller and greater values. But none of these families can be continued to the lower bandgap edge. The bifurcation points of solutions (d)-(f) are $\mu\approx-2.774$, $\mu\approx-2.765$ and $\mu\approx-2.777$, respectively. Our numerical finding here is that the families (d) and (f) are linearly stable for all values of $\mu$ up to some threshold $\mu^*\approx1.57$. Notice  that it is similar to the case of two in-phase FGSs. Moreover, in \cite{kizin} the equality of the stability thresholds among another families of gap solitons with codes $(+)$, $(++)$ and $(+++)$ was also observed. The existance of such threshold may be related to the FGS itself which is an elementrary entity for more complex gap solitons. For greater values of $\mu$, the families of two out-of-phase FGSs exhibit weak oscillatory instabilities similar to \cite{kizin}. The orders of the real parts of the unstable eigenvalues are approximately the same as in the case of two separated in-phase FGSs described above. The family (e) is unstable for the entire interval of its existence. 

Looking to these linear stability results, one can conjecture that the solutions from the second class {\it are linearly stable if two FGSs are separated by an odd number of empty potential wells} and the corresponding value $\mu$ is situated far enough from the upper bandgap edge. \textit{If the number of empty sites is even, then these solutions are supposed to be unstable.}

To study the dependence between instability of a gap soliton and the number of empty potential wells in its middle, we computed the spectra of five solutions with codes $(+0+)$, $(+00-)$, $(+000+)$, $(+0000-)$ and $(+00000+)$. All these gap solitons exhibit a strong exponential instability caused by a pair of real eigenvalues. In Figure~\ref{fig:distance} one can find the results of this experiment. Real parts of unstable eigenvalues decay exponentially as the number of empty potential wells grows. It may be related to the coupling and uncoupling of two separated FGSs due to short and long distance of separating, respectively. If the separated FGSs are uncoupled, then the corresponding gap soliton is linearly stable.

\begin{figure}\centering
\begin{psfrags}
\psfrag{A}[bl][bl][1.5]{$t$}
\psfrag{B}[bl][bl][1.5]{$t$}
\psfrag{C}[bl][bl][1.5]{$t$}
\psfrag{D}[bl][bl][1.5]{$x$}
\psfrag{E}[bl][bl][1.5]{$x$}
\psfrag{F}[bl][bl][1.5]{$x$}
\psfrag{G}[bl][bl][1]{$0$}
\psfrag{H}[bl][bl][1]{$20000$}
\psfrag{I}[bl][bl][1]{$40000$}
\psfrag{J}[bl][bl][1]{$0$}
\psfrag{K}[bl][bl][1]{$20000$}
\psfrag{L}[bl][bl][1]{$40000$}
\psfrag{M}[bl][bl][1]{$0$}
\psfrag{N}[bl][bl][1]{$250$}
\psfrag{O}[bl][bl][1]{$500$}
\psfrag{P}[bl][bl][1]{$-30$}
\psfrag{Q}[bl][bl][1]{$0$}
\psfrag{R}[bl][bl][1]{$30$}
\psfrag{S}[bl][bl][1]{$-30$}
\psfrag{T}[bl][bl][1]{$0$}
\psfrag{U}[bl][bl][1]{$30$}
\psfrag{V}[bl][bl][1]{$-30$}
\psfrag{W}[bl][bl][1]{$0$}
\psfrag{X}[bl][bl][1]{$30$}
\psfrag{Y}[bl][bl][1.2]{\textcolor{white}{(a)}}
\psfrag{Z}[bl][bl][1.2]{\textcolor{white}{(b)}}
\psfrag{a}[bl][bl][1.2]{\textcolor{white}{(c)}}
\includegraphics[width=0.75\textwidth]{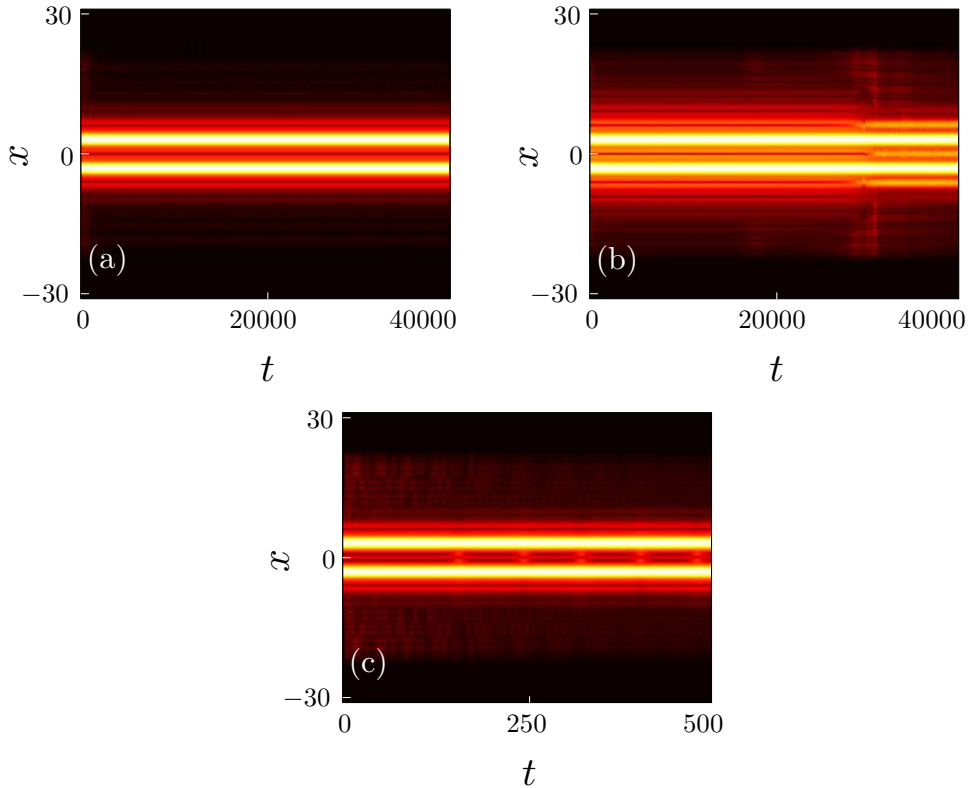}
\caption{Numerical integration of equation~(\ref{eq:1d}) with initial conditions depicted in Figure~\ref{fig:exam}(a) and (e). Pseudocolor plots of $|U(x,t)|^2$ are shown. Panels (a) and (b): the initial profile of the mode $(+0-)$ with parameter $\mu=1$ and $\mu=2$, respectively. Panel (c): the initial profile of the mode $(+0+)$ with parameter $\mu=1$. The spatial step $h\approx 0.1$ and the time step $\tau=0.01$. The initial profiles were perturbed by 3-percent perturbation.  The absorbing layers for $|x|>20$ were included to remove the reflection from the boundaries.}
\label{fig:evol}
\end{psfrags}
\end{figure}

\subsection{The dynamics of time-dependent GPE}

In order to check the predictions of linear stability analysis, we integrated numerically the time-dependent NLSE (\ref{eq:1d}) with initial conditions depicted in Figure~\ref{fig:exam}(a) and (d) with different parameters $\mu$ and $V_0$. As we mentioned above, we used a finite difference scheme with spatial step $h\approx 0.1$ and time step $\tau= 0.01$ in the domain $[-10\pi; 10\pi]\times[0; 40000]$. In order to eliminate the reflection from the boundaries, we added artificial absorbing layers for $x>20$ and $x<-20$. Besides, we perturbed the initial condition by means of $3$-percent perturbation, i.e. $U(x,0)\mapsto 1.03 \times U(x,0)$. The results of this experiment are shown in Figure~\ref{fig:evol}. Panels (a) and (b) correspond to mode $(+0-)$ for parameter $\mu=1$ and $\mu=2$, respectively. One can see that the results of numerical integration of (\ref{eq:1d}) are in a good agreement with linear stability analysis considered in the previous subsections. The gap soliton computed in the panel (a) preserved its shape for the whole time of computation while that in the panel (b) has been deformed due to weak oscillatory instabilities. Panel (c) exhibits the evolution of the mode $(+0+)$ for $\mu=1$. It has been confirmed by the simulations that this mode is unstable. Notice that the characteristic time necessary for the growing perturbation to manisfest itself is much smaller for the exponential instability than for the weak oscillatory instability.

One more observation is as follows. The temporal evolution of the unstable mode $(+0+)$ results {\it not} in its destroying, but in forming of some pulsating formation resembling a breather.  Simulations for the codes $(+00-)$, $(+000+)$ and different parameters $\mu$ and $V_0$ yield the similar result. We conjecture that this phenomenon is typical and 
{\it every exponentially unstable bound state, e.g. one  with the code $(+0\ldots0+)$, $(-0\ldots0-)$, $(+0\ldots0-)$ and $(-0\ldots0+)$, transforms into a pulsating  formation instead of being completely destroyed.}

\section{Conclusions}
\label{sec:discuss}

In the paper, we studied the stability of  gap solitons for the one-dimentional nonlinear Shr$\ddot{\rm o}$dinger equation (NLSE) with cosine potential  (\ref{eq:1d}).  It is known, that under certain conditions \cite{alfimov_2} each gap soliton of this equation can be coded by means of a bi-infinite sequence of symbols of some finite alphabet. In out study, the parameters $\mu$ and $V_0$ ($\mu$ is the chemical potential, see the anzats (\ref{eq:anzatz}), $V_0$ is the depth of the potential)  were chosen from the Region 1 in Figure~\ref{fig:bandgap} where three-symbols alphabet, ``$-$'', ``$0$'' and  ``$+$'', for the coding is relevant.
The gap solitons of interest are the bound states of a pair of fundamental gap solitons (FGSs) separated by some number of periods of the potential. In-phase bound states correspond to the codes $(+0\ldots 0+)$ and $(-0\ldots0-)$ (the top row in Figure~\ref{fig:exam}) and out-of-phase ones have the codes $(+0\ldots 0-)$ and $(-0\ldots0+)$ (the bottom row of Figure~\ref{fig:exam}).

In order to study linear stability of above-mentioned classes of gap solitons, we solved numerically the spectral problem (\ref{eq:spectral}) using the Fourier collocation method (FCM) and the Evans function method (EFM). It follows from our computations that some of the considered gap solitons are stable, some of them exhibit strong (exponential) instabilities and other ones undergo weak oscillatory instabilities. The modes with the codes $(+0-)$, $(+00+)$ and $(+000-)$ are linearly stable for all values of $\mu$ below a certain threshold. As $\mu$ exceeds this threshold, these modes exhibit weak oscillatory instabilities. On the contrary, the gap solitons with the codes $(+0+)$, $(+00-)$ and $(+000+)$ turned out to be unstable due to the presence of a pair of real eigenvalues in their spectra for all considered parameters $\mu$. As the result, we conjecture that two separated in-phase (out-of-phase) FGSs are linearly stable (unstable) for an even (odd) number of empty potential wells between them. Also, two in-phase (out-of-phase) FGSs are linearly unstable (stable) for an odd (even) number of empty sites between them.
The direct integration of (\ref{eq:1d}) by means of a finite-difference scheme supported the linear stability results. While all the results presented above correspond to $V_0=6$, we have also checked that the main conclusions of our study  remain valid for several other values of $V_0$ (such as $V_0=5$ and $V_0=7$).

It is interesting that the results of this study confirm the known results on stability/instability of intrinsic localized modes (ILM) for the discrete nonlinear Schr$\ddot{\rm o}$dinger equation (DNLSE). It is known that the DNLSE can be regarded as an approximation for the  NLSE with periodic potential and ILM can be used for qualitative description of gap solitons in the first bandgap \cite{AKKS,FS14} (so-called tight-binding limit). Rigorous statement about the number of unstable eigenvalues for ILM can be found in \cite{KevPel}. Theorem 3.6 of \cite{KevPel} connects  the number of unstable eigenvalues for an ILM with the number of flips in its code. In \cite{KevPel} this result was formulated for the case of the DNLSE with focusing nonlinearity and is valid near so-called anticontinuous limit only. Translating this statement to case of the defocusing DNLSE can be made by means of the staggering transformation. However, using the analogy between the DNLSE and the periodic NLSE, one can expect instability of the bound states for the codes $(+0\ldots 0+)$ and $(-0\ldots 0-)$ with odd number of zero symbols between nonzero symbols and for $(+0\ldots 0-)$ and $(+0\ldots 0-)$ with even number of zero symbols, exactly as it is claimed in this paper.  At the same time, the analogy with the discrete model fails to reveal quite delicate oscillatory instabilities described above that were not found in the case of ILM.

A further generalization of this study may be associated with the following question: {\it does there exist  a general relation between the stability of a gap soliton and its code?} We believe that the numerical experience summarized in \cite{yang} and in the present paper may be extended to gap solitons with more complex codes, as well as to the gap solitons from higher bandgaps. However, this interesting issue lies beyond the scope of this paper.

\section{Acknowledgment}
Author is grateful to Prof. Georgy L. Alfimov and Dr. Dmitry A. Zezyulin for a careful reading of the manuscript, significant remarks and discussions.

\end{document}